\documentstyle[float,epsfig,aps]{revtex}

\def\bfi{\begin{figure}}
\def\efi{\end{figure}}
\def\fref#1{Fig.~\ref{#1}}

\preprint{NUC-MINN-99/3-T, February 1999}

\begin{document}


\title{Imaging the Space-Time Evolution of High Energy Nucleus-Nucleus
Collisions with Bremsstrahlung}

\author{J.I. Kapusta\footnote{kapusta@physics.spa.umn.edu}
and S.M.H. Wong\footnote{swong@nucth1.hep.umn.edu}}

\address{
School of Physics and Astronomy,
University of Minnesota,
Minneapolis, MN 55455}

\date{March 2, 1999}

\maketitle

\begin{abstract}
The bremsstrahlung produced when heavy nuclei collide is estimated
for central collisions at the Relativistic Heavy Ion Collider.
Bremsstrahlung photons with energies below 100 to 200 MeV are
sufficient to discern the gross features of the space-time evolution
of electric charge, if they can be separated from other sources of
photons experimentally.  This is illustrated explicitly by considering
two very different models, one Bjorken-like, the other Landau-like,
both of which are constructed to give the same final charge rapidity
distribution.
\end{abstract}

\null
\hspace{1.5cm}{PACS numbers: 25.75.-q, 13.40.-f}
\hspace{4.5cm}{preprint NUC-MINN-99/3-T}\\

\section{Introduction}

The Relativistic Heavy Ion Collider (RHIC) at Brookhaven National
Laboratory (BNL) will have its first beam available for experiments
by the end of 1999.  A basic quantity of interest is the initial
energy density of hot matter whose study is the ultimate physics goal.
Hadronic measurements may not illuminate the state of the system at maximum
density for two reasons: First, the system must expand and cool to a
sufficiently low density before quarks and gluons can bind into independent
color singlet hadrons.  Second, even after hadronization or confinement phase
transition the hadrons can scatter among themselves, changing their momentum
distributions.  For these reasons hard photons and lepton pairs have been
considered better probes of the high temperature plasma of quarks and gluons
since the production rate of such radiation increases dramatically
with temperature.  However, high transverse momentum photons and lepton pairs
can also originate from the lower temperature
hadronic phase due to hadron-hadron collisions and decays
coupled with collective transverse expansion.
(Focussing on high mass pairs negates this to some extent but eventually the
Drell-Yan mechanism will overwhelm these at very high mass.)  If it was only
possible to see these nuclear collisions with the human eye!  This is not
possible, of course, leaving the next best option: Measurement of the
electromagnetic bremsstrahlung produced by the massive deceleration of the
nuclear electric charge.

The first theoretical study of nucleus-nucleus bremsstrahlung at relativistic
energies, of order 1 GeV per nucleon, was carried out by one of the authors
\cite{kap}.  This was followed by Bjorken and McLerran who were interested
in much higher beam energies \cite{BM}, and then by Dumitru, McLerran and
St\"{o}cker \cite {Dumitru}.  Most recently Jeon, Kapusta, Chikanian and
Sandweiss \cite{gangof4} studied bremsstrahlung photons with energies less than
5 MeV and a detector design to measure them at RHIC.  Such an observation would
allow for a determination of the final rapidity distribution of the net electric
charge, which is one measure of transparency and stopping, obviously important
for the initial energy density and the formation of quark-gluon plasma.  It may
be argued that a direct observation of all electrically charged hadrons would
yield the same information, albeit at much higher cost.  Nevertheless a cross
check would be valuable.

In this paper, we wish to determine whether it is possible to discern the gross
features of the space-time evolution of the electric charge using bremsstrahlung
photons with energies up to 100 or 200 MeV in the lab frame of RHIC (which is
also the center-of-mass frame for a collider).  Generally a photon of energy 200
MeV would allow one to probe distances on the scale of 1 fm and times on the
scale of 1 fm/c.  At ultrarelativistic beam energies this is not quite true due
to the extreme forward focussing of the bremsstrahlung.  Nevertheless the answer
to the question posed at the beginning of this paragraph is yes.  In section 2,
we work out the necessary formulas for bremsstrahlung.  In section 3, we apply
it
to two extreme models: the first is Bjorken-like, which has considerable nuclear
transparency, and the second is Landau-like, which has maximum nuclear stopping.
Both models are tuned to produce a flat rapidity distribution for the net
electric charge so that measurement of hadrons alone could not distinguish them.
Section 4 contains concluding remarks.

\section{Calculation of Bremsstrahlung}

Consider a central collision of two equal mass nuclei of charge $Z$
in their center of momentum frame.  Denote the speed of each beam by
$v_0$ and the corresponding rapidity by $y_0$, so that the
rapidity gap between projectile and target nuclei is 2$y_0$.
Velocity and rapidity are related by $v = \tanh y$.
Before collision each nucleus may be viewed as a flattened pancake
with negligible thickness since the photons we are considering cannot probe
distances smaller than 1 fm.  For example, with a RHIC beam energy of 100 GeV
per nucleon the Lorentz $\gamma$ = 106 so that the diameter of a gold nucleus is
contracted to 0.13 fm!

The formula for computing the classical bremsstrahlung from accelerated charges
is well-known \cite{Jackson}.  For charges $q_i$ with coordinates
${\bf r}_i(t)$,
velocities ${\bf v}_i(t)$ and accelerations ${\bf a}_i(t)$ the intensity and
number of photons emitted with frequency $\omega$ in the direction ${\bf n}$ are
\begin{equation}
\frac{d^2I}{d\omega d\Omega} = \omega \frac{d^2N}{d\omega d\Omega} = |{\bf A}|^2
\end{equation}
where the amplitude is
\begin{equation}
{\bf A} = \sum_i q_i \int_{-\infty}^{\infty} \frac{dt}{2\pi}
\exp\left\{ i\omega \left( t-{\bf n}\cdot {\bf r}_i(t) \right) \right\}
\frac{{\bf n}\times \left[\left( {\bf n}-{\bf v}_i(t) \right) \times
{\bf a}_i(t) \right]}{\left( 1 - {\bf n} \cdot {\bf v}_i(t) \right)^2} \, .
\end{equation}
The sum over charges may of course be replaced by an integral when the charge
distribution is viewed as continuous.

\subsection{Soft bremsstrahlung}

For low frequency photons, the nucleus-nucleus collision appears
almost instantaneous.  To the extent that the transverse rapidities
of the outgoing charged particles are small compared to the beam
rapidity one may then derive the formula \cite{gangof4}:
\begin{equation}
\frac{d^2I}{d\omega d\Omega} = \frac{\alpha}{4\pi^2}
\sin^2\theta \left| \int d^2r_{\perp} \sigma(r_{\perp})
e^{-i \omega {\bf n}\cdot {\bf r}_{\perp}}
\left[ \int dy \frac{v(y) \rho(r_{\perp},y)}
{1-v(y)\cos\theta} - \frac{2v_0^2 \cos\theta}{1-v_0^2
\cos^2\theta} \right] \right| ^2 \, .
\end{equation}
Here it is convenient to take ${\bf n} = \left(\sin\theta, 0, \cos\theta
\right)$.  The function $\rho$ represents the charge rapidity distribution at a
distance $r_{\perp}$ from the beam axis $z$ and is normalized as:
\begin{equation}
\int_{-\infty}^{\infty} dy \rho(r_{\perp},y) = 2 \, ,
\end{equation}
the 2 arising because the total charge is 2$Z$.
We cannot go further without some knowledge of the distribution
$\rho(r_{\perp},y)$.

It is advantageous to have a simple analytic model with a
variable charge rapidity distribution.  To this end, suppose that
$\rho(r_{\perp},y)$ is independent of $r_{\perp}$.  Then
\begin{equation}
\frac{d^2I}{d\omega d\Omega} = \frac{\alpha Z^2}{4\pi^2}
\sin^2\theta \left| F(\omega \sin\theta) \right|^2 \left|
\left[ \int dy \frac{v(y) \rho(y)}
{1-v(y)\cos\theta} - \frac{2v_0^2 \cos\theta}{1-v_0^2
\cos^2\theta} \right] \right| ^2 \, ,
\end{equation}
where $F$ is a transverse nuclear form factor:
\begin{equation}
F = \frac{1}{Z} \int d^2r_{\perp} \, \sigma\left( r_{\perp} \right)
e^{-i \omega {\bf n}\cdot {\bf r}_{\perp}}\, ,
\end{equation}
where $\sigma$ is the charge per unit area of the nucleus.  A solid sphere
approximation should be adequate for large nuclei,
in which case
\begin{equation}
F(q) = \frac{3}{q^2}\left( \frac{\sin q}{q} - \cos q \right) \, ,
\end{equation}
where $q=\omega R \sin\theta$ and $R$ is the nuclear radius.
For small angles and low frequencies the nuclear form factor is practically
equal to one.

The integral over rapidity can easily be performed for
a flat rapidity distribution: $\rho(y) = 1/y_0$ for $-y_0 < y < y_0$.
Then the photon distribution is:
\begin{eqnarray}
\frac{d^2I}{d\omega d\Omega} &=& \frac{\alpha Z^2}{4\pi^2}
\sin^2 \theta \left| F(\omega R \sin\theta) \right|^2 \nonumber \\
&\times& \left[ \frac{2\cos\theta}{\sin^2 \theta} -
\frac{1}{y_0 \sin^2 \theta}
\ln\left( \frac{1+v_0 \cos\theta}
{1-v_0 \cos\theta} \right) - \frac{2v_0^2 \cos\theta}{1-v_0^2
\cos^2\theta} \right] ^2 \, .
\end{eqnarray}
Because $v_0$ is close to 1 the distribution deviates strongly from
the quadrupole form, peaking near small but nonzero
$\theta$.  At RHIC the peak occurs at $\theta \sim 1^{\circ}$.  The distribution
is plotted in \fref{f:f1} for central gold collisions at RHIC with
$\omega << 1/R = 30$ MeV.
\bfi
\centerline{
\epsfig{figure=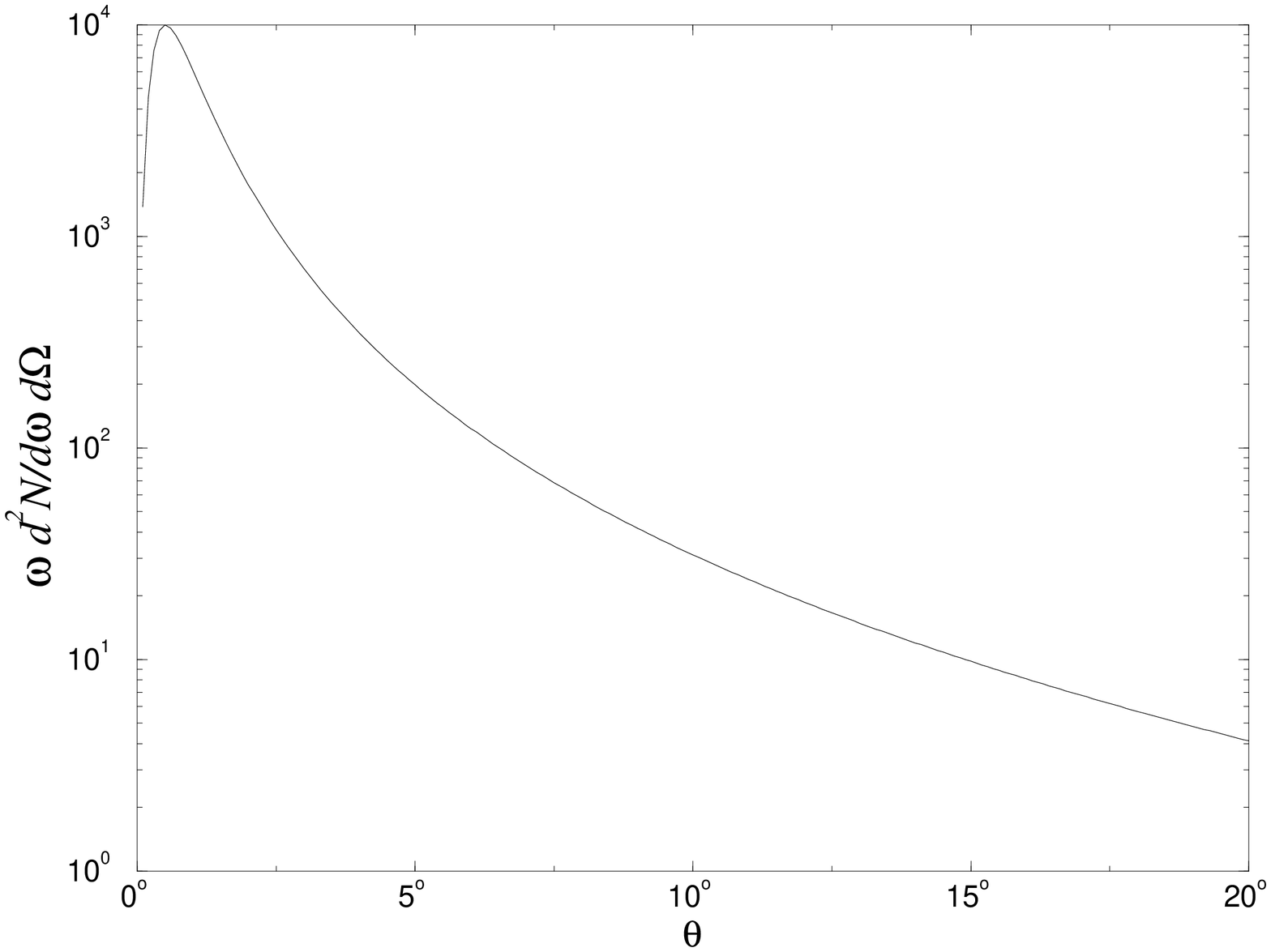,width=3.0in,angle=270}
}
\caption{The intensity as a function of angle for very soft photons which
results from a flat charge rapidity distribution.  The intensity is independent
of frequency for $\omega \ll 1/R$.}
\label{f:f1}
\efi

\subsection{Hard bremsstrahlung}

All the results above were already reported in \cite{gangof4} and are applicable
for very soft photons.  Here we will consider photons with energies up to 100 to
200 MeV with the expectation that they will reveal the spatial and temporal
development of the nuclear collision.  We will consider only central collisions
which can be triggered on experimentally with hadronic detectors, such as a
calorimeter.

For this purpose one can certainly invoke specific space-time models of nuclear
collisions at RHIC.  In this paper we will consider simple parameterizations of
such collisions which represent extreme limits of such models.  The general
formulas will be worked out here, in this section, and then applied to these
models in the next
section.

We will represent a large nucleus, such as gold, by a uniform spherical
distribution of nucleons with proper density $n_0 = 0.155$ nucleons/fm$^3$.  An
infinitesimal unit of charge will have a coordinate specified by $r_{\perp}$ and
by the distance $l$ parallel to the $z$-axis as measured from the backside of
the nucleus.  For given $r_{\perp}$ the allowed value of $l$ ranges from 0 to
$l_{max}(r_{\perp}) = 2\sqrt{R^2- r_{\perp}^2}$.  Summation over all
charges $i$ in eq. (2) is then replaced by an integral over ${\bf r}_{\perp}$
and $l$.  See \fref{f:f2}.

Each charge element undergoes some deceleration to a final rapidity y.  This
deceleration will take a finite time and can vary from one element to another.
In this paper we will fix the final observed charge rapidity distribution and
vary the space-time evolution to see whether hard bremsstrahlung can provide
information not obtainable by measuring hadrons only.  For definiteness we shall
fix the final charge rapidity distribution to be flat.  In individual
nucleon--nucleon
collisions it is known to be approximately proportional to $\cosh y$ in the
center-of-momentum frame \cite{lexus}.  In order to obtain a flat rapidity
distribution in a central nucleus-nucleus collision different charge elements
must have a rapidity that depends on $r_{\perp}$.  The assumed relationship
among these variables is
\begin{equation}
l = f(r_{\perp})y + g(r_{\perp})\sinh y
\end{equation}
where $f$ and $g$ are functions yet to be specified.  This means that the last
charge element in line will eventually come to rest in the center-of-momentum
frame.
\bfi
\centerline{
\epsfig{figure=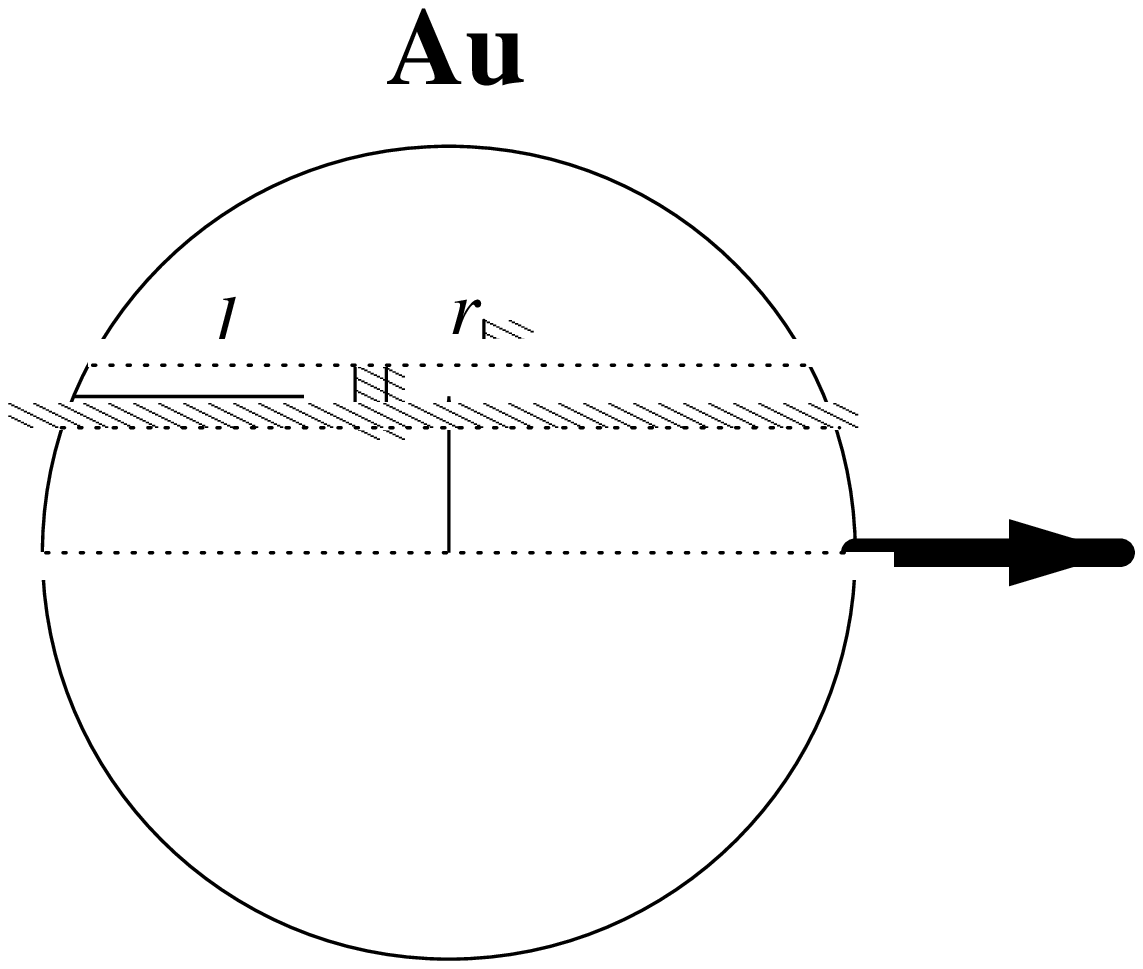,width=2.0in}
}
\caption{The charge element coordinate labels in the rest frame of the
nucleus.}
\label{f:f2}
\efi

The charge per unit area per unit rapidity is:
\begin{equation}
\frac{d^3Q}{d^2r_{\perp}dy} = n_0 \frac{Z{\rm e}}{A} \frac{dl(r_{\perp},y)}{dy}
= n_0 \frac{Z{\rm e}}{A} \left[f(r_{\perp}) + g(r_{\perp})\cosh y \right] \, .
\end{equation}
To have a net flat rapidity distribution requires that
\begin{equation}
\int_0^R dr_{\perp} r_{\perp} g(r_{\perp}) = 0 \, .
\end{equation}
As $r_{\perp} \rightarrow R$ the distribution should have a minimum at $y = 0$,
in accord with proton-proton scattering, whereas when $r_{\perp} \rightarrow 0$
the distribution must have a maximum at $y = 0$ to produce a distribution that
is flat overall.  This is physically sensible because there ought to be
significantly more deceleration in the central core of the nucleus than at its
outer periphery.  A linear extrapolation of nucleon-nucleon scattering to
central collisions of gold nuclei at RHIC \cite{gangof4,lexus} produces just
such a behavior with a net distribution that is approximately flat in rapidity.
We choose the parameterization $g(r_{\perp}) = g_1 l_{max}(r_{\perp}) - g_2
l^2_{max}(r_{\perp})$ where $g_1$ and $g_2$ are positive constants.  The
requirement of a net rapidity distribution which is flat results in $g_2 = 2
g_1/3R$.  The physically sensible requirement that the leading charge elements
emerge from the collision with the original beam rapidity determines the
function $f(r_{\perp})$ via $l_{max}(r_{\perp}) = f(r_{\perp})y_0 +
g(r_{\perp})\sinh y_0$.  Finally, the requirement that the charge rapidity
distribution be non--negative for all allowable $y$ leads to the condition $g_1
\le 3/(y_0 \cosh y_0 - \sinh y_0)$.  For definiteness we choose the upper limit
$g_1 = 3/(y_0 \cosh y_0 - \sinh y_0)$  since this value brings the rapidity
distribution as close to $\cosh y$ as $r_{\perp} \rightarrow R$ as possible,
consistent with free space nucleon-nucleon scattering.  The resulting
$d^3Q/d^2r_{\perp}dy$ is plotted in \fref{f:f3}, and is very
similar to what is obtained in LEXUS (see Fig. 1 of \cite{gangof4}).
\bfi
\centerline{
\epsfig{figure=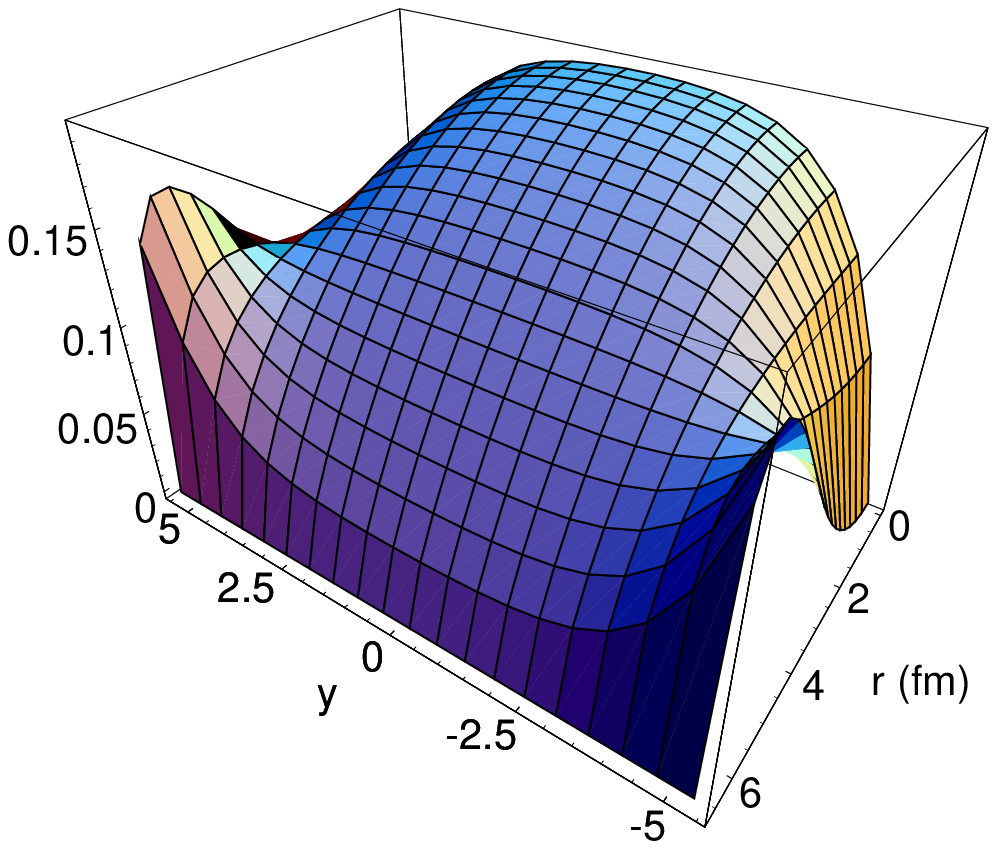,width=2.0in}
}
\caption{The charge rapidity distribution, $d^3 Q/d^2 r_{\perp} dy$, as it depends 
on the transverse coordinate $r_{\perp}$.}
\label{f:f3}
\efi

For a central collision of equal mass nuclei at RHIC the primary accelerations
occur along the direction of the beam axis.  By azimuthal symmetry the
transverse accelerations will, in addition, tend to cancel in the expression for
the amplitude.  The contribution to the amplitude from the projectile will then
be:
\begin{eqnarray}
A_+ &=& \sin\theta \frac{Z{\rm e}}{A} n_0 \int_0^R dr_{\perp} r_{\perp}
J_0(r_{\perp}\omega \sin\theta) \int_0^{l_{max}(r_{\perp})} dl \nonumber  \\
&\times& \int_{-\infty}^{\infty} dt \exp\{ i\omega\left( t- z(l,t)\cos\theta
\right)\} \frac{a(l,t)}{(1 - v(l,t)
\cos\theta)^2} \, .
\end{eqnarray}
The expression for the contribution from the target nucleus $A_-$ will be the
same except for the reversal of sign of the position, velocity, and
acceleration.  Since
there is a one-to-one relation between the longitudinal label $l$ and the
rapidity $y$ this amplitude may be written as an integral over $y$ instead of
over $l$ as:
\begin{eqnarray}
A_+ &=& \sin\theta \frac{Z{\rm e}}{A} n_0 \int_0^R dr_{\perp} r_{\perp}
J_0(r_{\perp}\omega \sin\theta) \int_0^{y_0} dy \frac{dl(y,r_{\perp})}{dy}
\nonumber \\
&\times& \int_{-\infty}^{\infty} dt \exp\{i\omega \left(t - z(y,t)\cos\theta
\right)\} \frac{a(y,t)}{(1 -
v(y,t) \cos\theta)^2} \, .
\end{eqnarray}
For a specified model of the deceleration this three-dimensional integral must
be evaluated numerically.

\section{Two Extreme Models}

In this section we will describe two extreme models for the space-time evolution
of a central nucleus-nucleus collision at RHIC.  The real world will undoubtedly
be different than either of these.  However, our point is that hard
bremsstrahlung may be used to discern the gross features of the evolving charged
matter, something that hadron detection alone cannot do because the final charge
rapidity distribution in these two models is the same (by construction).

\subsection{A Bjorken-like model}

This model assumes that the nucleons undergo a constant deceleration upon
nuclear overlap.  It essentially assumes transparency \cite{Bj} of electric
charge although the final charge rapidity distribution is assumed to be flat.
Because the nuclei are so highly Lorentz contracted initially it is an excellent
approximation to assume that all charge elements are located at $z = 0$ at the
moment $t = 0$ of complete nuclear overlap.  A photon with energy less than 200
MeV would never be able to distinguish this from a Lorentz contracted pancake of
thickness 0.13 fm!  For a projectile charge element with nuclear coordinates
$l, r_{\perp}$ the position as a function of time is:
\begin{eqnarray}
{z(l,r_{\perp},t) \, = \, \left\{ \begin{array}{ll}
v_0 t & t < 0 \\
v_0 t + a(l,r_{\perp})t^2/2 & 0 < t < t_f \\
z(l,r_{\perp},t_f) + v_f (t - t_f) & t_f < t \, .
\end{array} \right. }
\end{eqnarray}
The deceleration is
\begin{equation}
a(l,r_{\perp}) = \frac{v_f - v_0}{t_f} = \frac{\tanh y - \tanh
y_0}{t_f} \, .
\end{equation}
Remember that $l$ and $v_f = \tanh y$ are related by eq. (9).  The
trajectory of
a target charge element is obtained symmetrically.  Note that the charges
continue to decelerate even after the nuclei have passed through each other.
This may be thought of as strings connecting projectile and target charges
causing them to decelerate at a distance, or by the finite time necessary
for particle production.

The deceleration time $t_f$ is allowed to depend on the length of the tube
as follows:
\begin{equation}
t_f(r_{\perp}) = t_0 + l_{max}(r_{\perp})/c_0
\end{equation}
where $t_0$ and $c_0$ are constants.  Several choices of these parameters will
be considered.

\subsection{A Landau-like model}

This model assumes that the nuclei stop each other within a distance of one
Lorentz contracted nuclear diameter, then accelerate longitudinally due to the
high pressure \cite{Landau}.  The deceleration occurs over such a short time
interval that photons of 200 MeV or less cannot resolve it.  Thereafter our
variation of Landau's model assumes that the charge elements accelerate
uniformly to their final velocities.  The trajectories are parameterized as:
\begin{equation}
a(l,r_{\perp}) = -v_0 \delta(t) + \frac{v_f}{t_f} \theta(t)
\theta(t_f-t) \, .
\end{equation}
The relationship among $l$, $v_f$ and $y$ is as above.  In a
hydrodynamical approach the final time $t_f$ is expected to be independent
of $r_{\perp}$ but it can be left as a parameter, just as in the Bjorken-like
model.

\subsection{Comparison of models}

Before comparing the bremsstrahlung produced in these two models let us compare
the space-time evolutions.  In \fref{f:f4} we plot cells bounded by lines of
constant $l$ and $r_{\perp}$ in the $x-z$ plane as a function of time.  Each
cell contains the same amount of charge.
The left panel shows the Bjorken-like model and
the right panel shows the Landau-like model.  Both models have $t_f =
1$ fm/c; that is, the acceleration time is independent of $r_{\perp}$.
There are several noteworthy features: The Bjorken-like model always
has a gap in the beam direction because it takes a finite time
for the charges to decelerate.  The charges that finally come to rest in this
frame of reference require 1 fm/c to decelerate from $v = \pm v_0$ to $v
= 0$ and by then they have traveled 0.5 fm.  In contrast, the charges in the
Landau-like model stop instantaneously and then accelerate to their final
velocities.  Therefore there is no gap in the beam direction.  Note that the
shaded regions indicate where the net charge is nonzero.  The gap in the
central region in the Bjorken-like model does not signify empty space but rather
a region where there is zero net charge; it may be occupied by
quark-gluon plasma or hadronic matter or strings or something else.
\bfi
\centerline{
\hbox{
\epsfig{figure=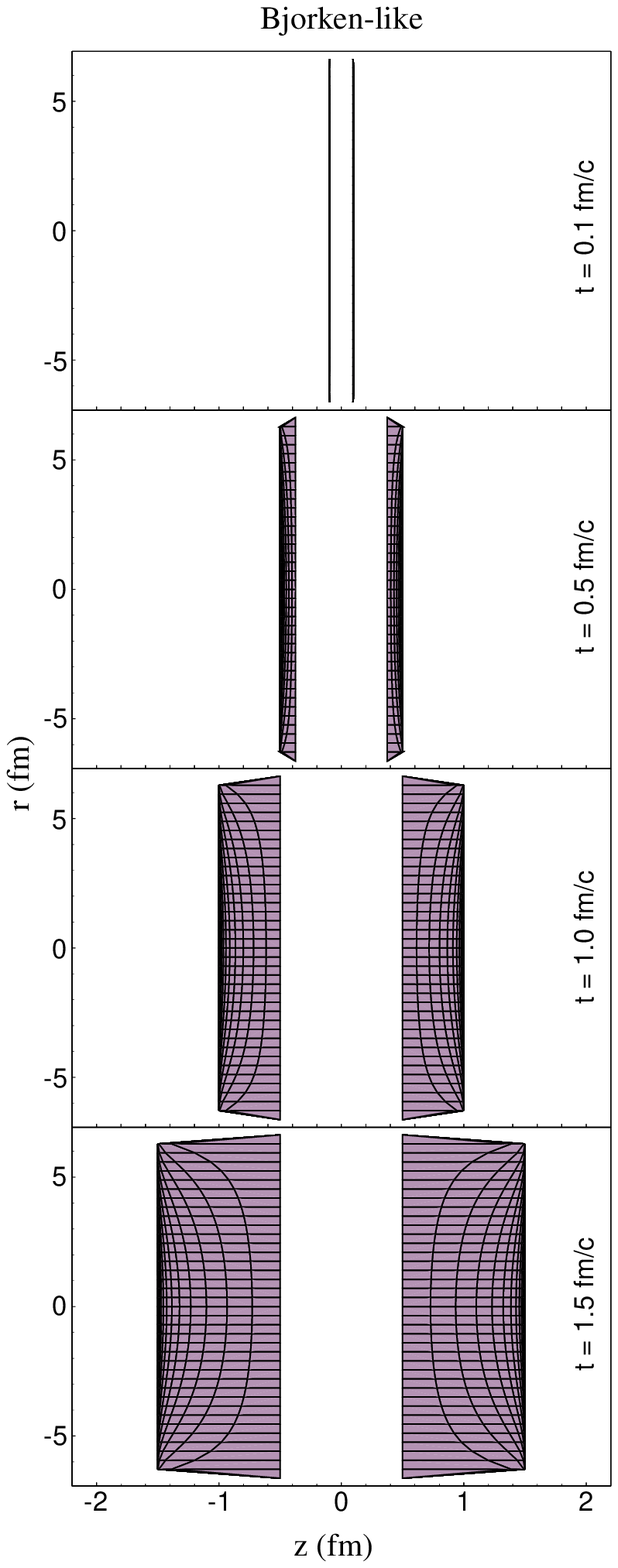,width=2.0in} \hspace{2.0cm}
\epsfig{figure=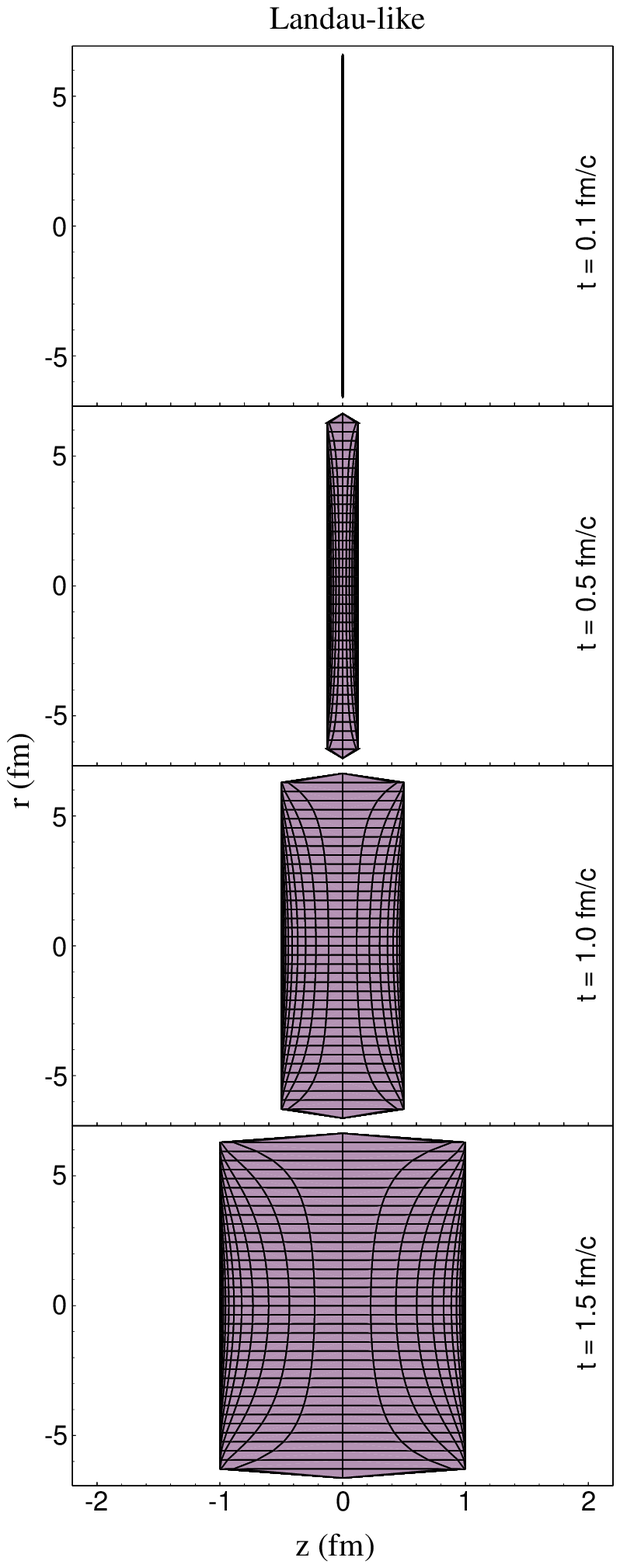,width=2.0in}
}}
\caption{ Space-time evolution in a Bjorken-like model compared to that of
a Landau-like model both with $t_f = 1$ fm/c.  The curves indicate contours
of constant $l$ and $r_{\perp}$ and so each cell, as delineated this way,
contains the same amount of electric charge.}
\label{f:f4}
\efi

In \fref{f:f5} we show the space-time evolution for $t_0 = 0.5$ fm/c and $c_0 = 1$
in the Bjorken-like model and for $t_0 = 10$ fm/c and $c_0 = \infty$ in the
Landau-like model.  We chose the acceleration time in the Landau-like model to
be independent of $r_{\perp}$ because this seems more relevant for a
hydrodynamical
model.  In the Bjorken-like model we chose the minimum time $t_0 = 0.5$ fm/c
because, even in free space, nucleon-nucleon collisions must occur over some
finite time interval.  The value of 0.5 fm/c is suggested by the analysis of
Gale, Jeon and Kapusta \cite{DY-J} of Drell-Yan and $J/\psi$ production in very
high energy proton-nucleus collisions.  The central cores of the gold nuclei
then require 13.9 fm/c to decelerate to their final velocities.  Hence the
average deceleration time ought to be comparable to the 10 fm/c chosen for
the Landau-like model.  Apart from the quantitative difference in the space-time
evolution, the gap in the Bjorken-like model is now more like an ellipse than a
slab.
\bfi
\centerline{
\hbox{
\epsfig{figure=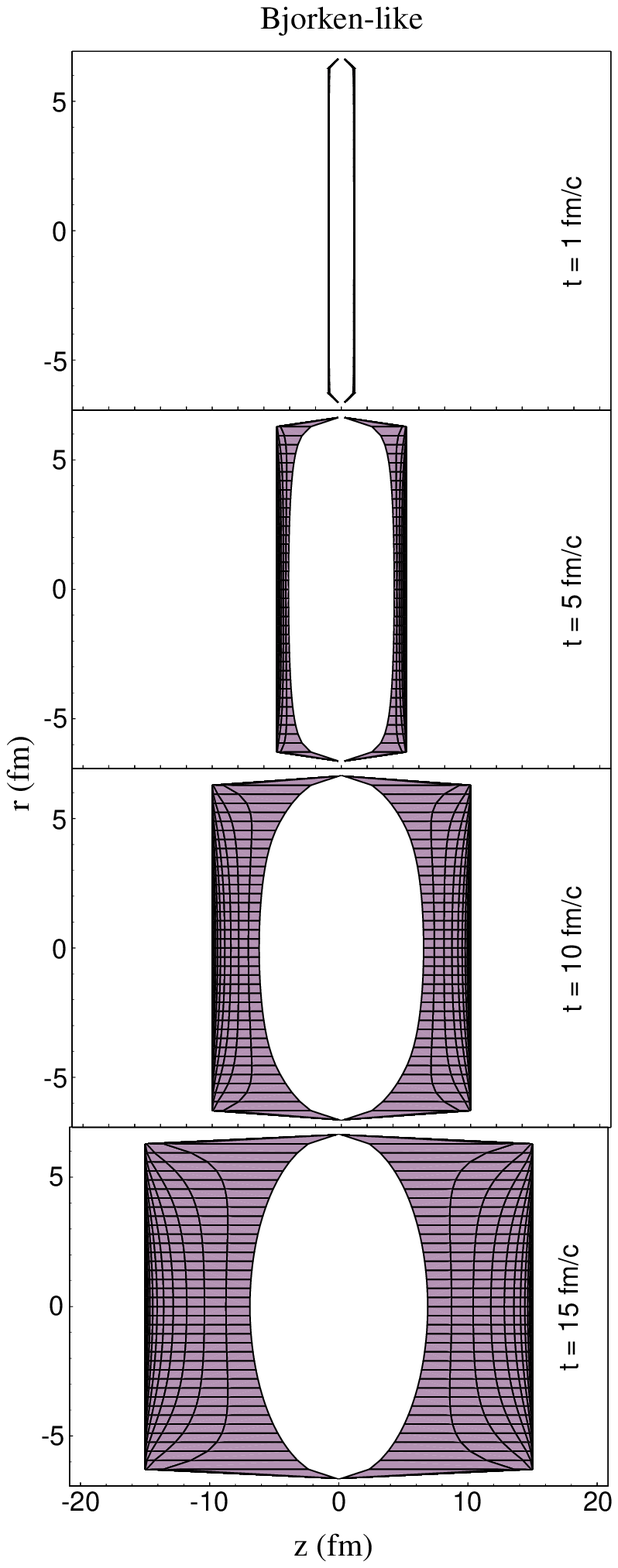,width=2.0in} \hspace{2.0cm}
\epsfig{figure=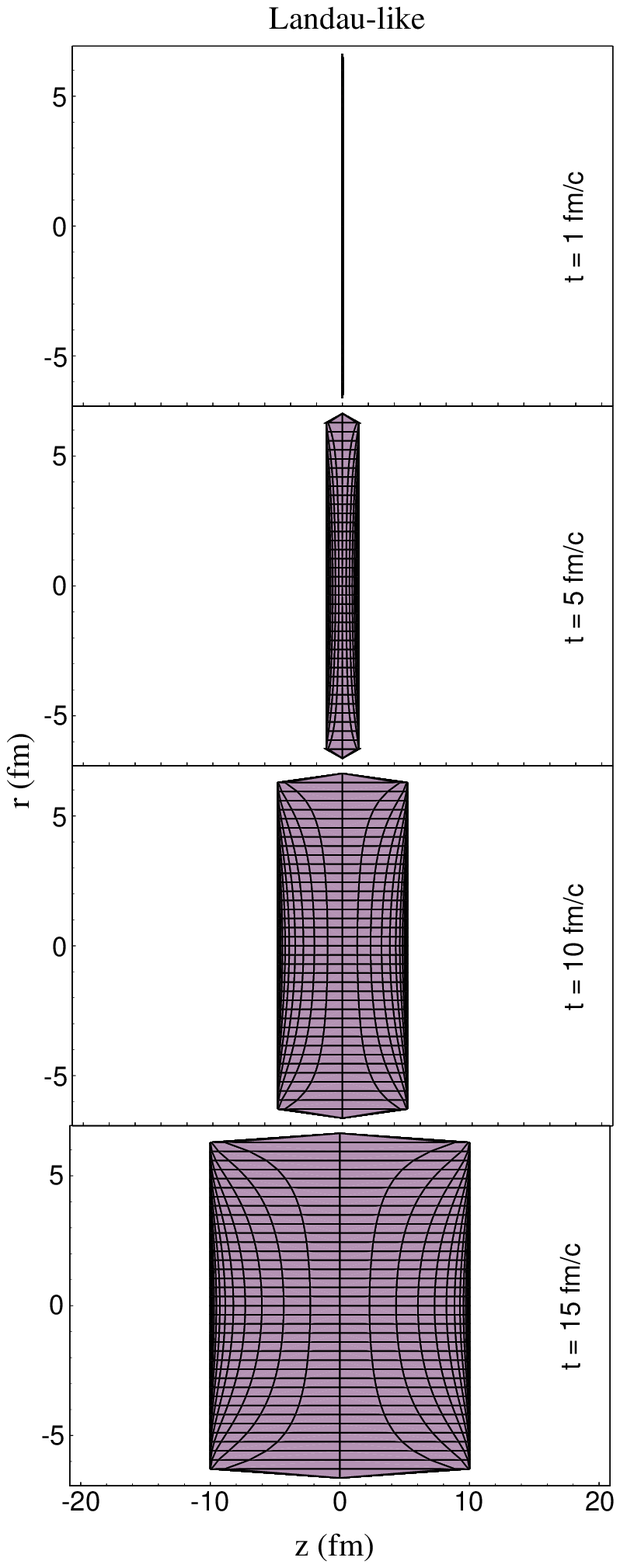,width=2.0in}
}}
\caption{Same as \fref{f:f4} except that the Bjorken-like model has
a $t_f$ varying from 0.5 fm/c at $r_{\perp} = R$ to 13.9 fm/c at
$r_{\perp}=0$, and the Landau-like model has fixed $t_f = 10$ fm/c.}
\label{f:f5}
\efi

The intensity distribution in the Bjorken-like model is plotted in \fref{f:f6} for
the angles 1$^o$, 5$^o$, 10$^o$ and 20$^o$.  Three values of $c_0$ have been
chosen for display: 0.5, 1 and $\infty$.  The value of $t_0$ has
been fixed at 0.5 fm/c in all cases, although the results are practically the
same with any $t_0 < 1$ fm/c.  The three values of $c_0$ correspond
to final times for the central cores of 27.3, 13.9 and 0.5 fm/c, respectively.
(Be careful to note the false zeroes in three of the four panels.)
The dependence on $c_0$ is very weak considering the wide range
of resulting $t_f$.  This may be understood as a relativistic
effect.  Because of the relation $v = \tanh y$, most of the outgoing
charges still move with a speed nearly equal to that of light.
Therefore the phase in eqs. (12-13) is approximately
$i \omega t(1-\cos\theta)$, and for it to become comparable to 1
requires an $\omega$ much bigger than naively expected.  The
separation of the curves increases with $\theta$ following this
phase factor.  The primary falloff of the intensity with increasing
photon energy is caused by the transverse form factor described
by eq. (7).  The falloff increases with angle because the argument
of the form factor is $q=\omega R \sin\theta$.

The intensity distribution in the Landau-like model is plotted in \fref{f:f7}
for the same set of angles.  Four illustrative values of $t_f$
have been chosen:  1, 2, 10 and 20 fm/c.  In this model there is a
striking dependence on $t_f$.  The oscillatory behavior for
large $t_f$ is due to the interference between the initial
instantaneous stopping with the subsequent acceleration to the final
velocities.  At 10$^o$ and with $t_f = 20$ fm/c, for example,
the enhancement due to constructive interference is an order of
magnitude at $\omega = 60$ MeV as compared to the soft photon limit.

\bfi
\centerline{
\epsfig{figure=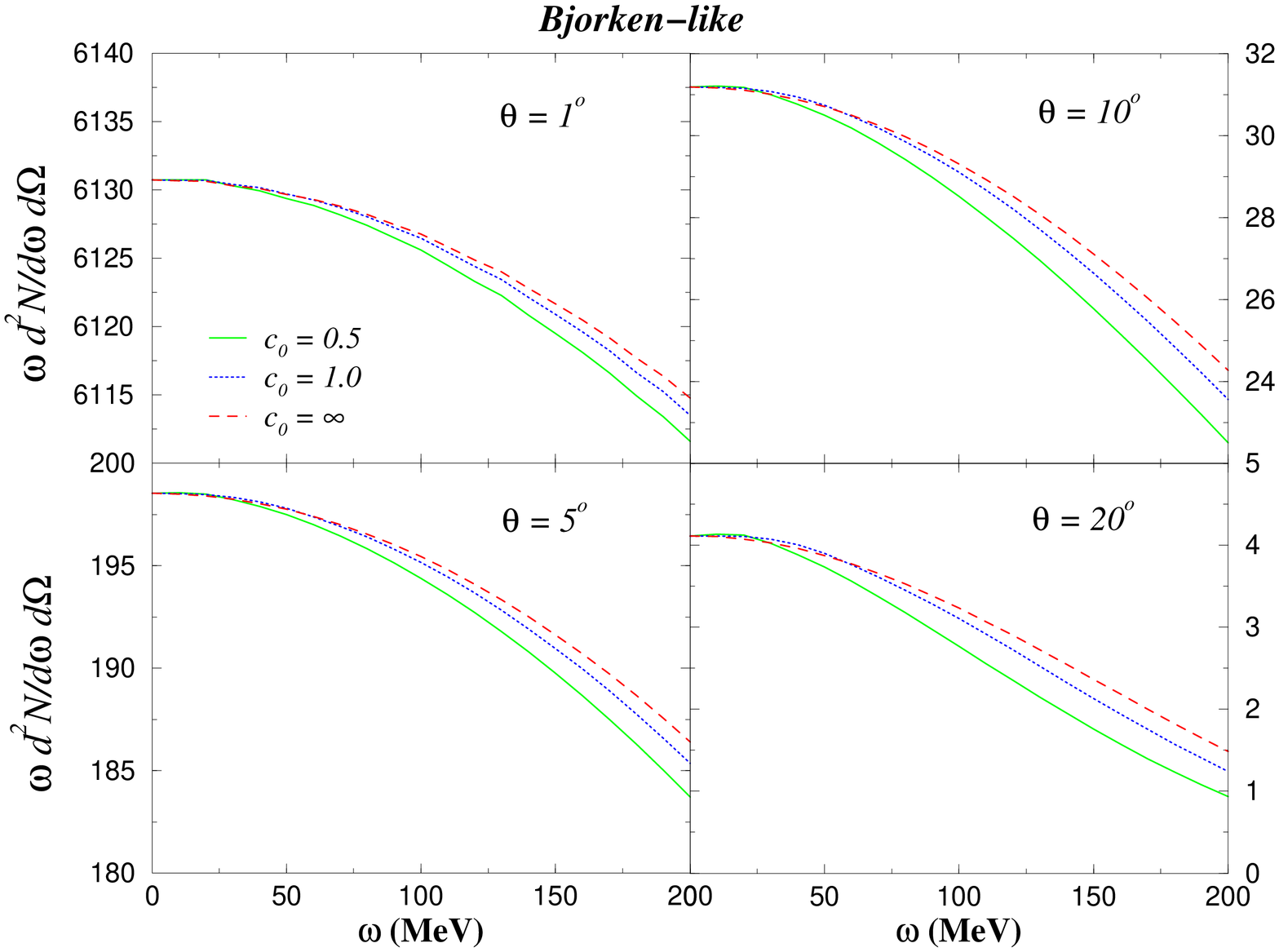,width=4.0in,angle=270}
}
\caption{The intensity distribution in the Bjorken-like model
as a function of photon energy for various angles.
The three curves correspond to different acceleration times.}
\label{f:f6}
\efi

\bfi
\centerline{
\epsfig{figure=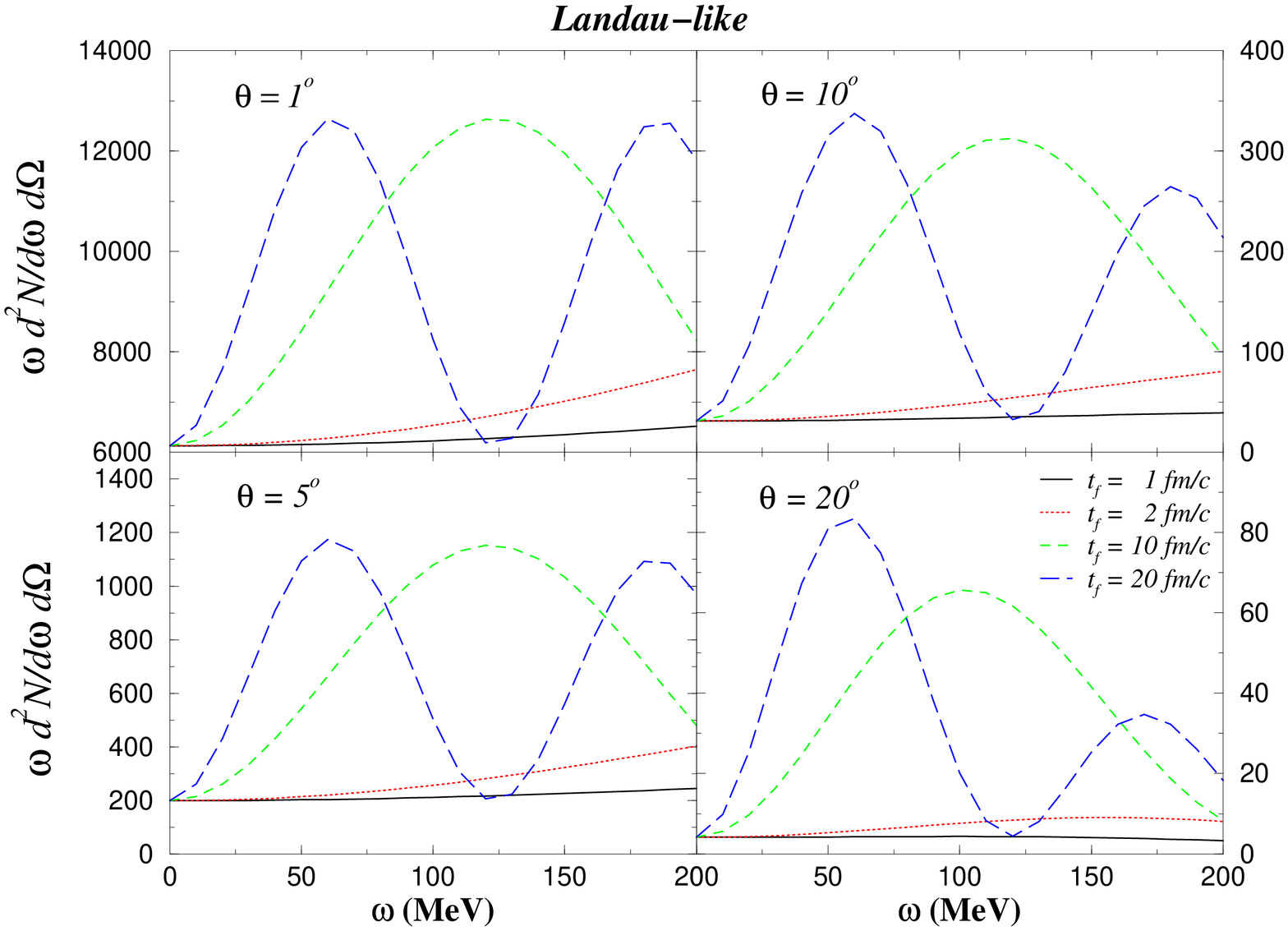,width=4.0in,angle=270}
}
\caption{ The intensity distribution in the Landau-like model
as a function of photon energy for various angles.
The four curves correspond to different acceleration times.}
\label{f:f7}
\efi

The oscillatory behavior which arises in the Landau-like model
but not in the Bjorken-like model is easy to understand in a simplified
version of these models.  Let us shrink each nucleus to a point charge.
In the Bjorken-like version, assume that the charges decelerate from
$\pm v_0$ to $\pm v_f$ instantaneously.  The intensity distribution
is:
\begin{equation}
\frac{d^2I}{d\omega d\Omega} = \frac{Z^2 \alpha}{\pi^2}
\sin^2\theta \cos^2\theta \frac{(v_0^2-v_f^2)^2}
{(1-v_0^2\cos^2\theta)^2 (1-v_f^2\cos^2\theta)^2} \, ,
\end{equation}
which has no frequency dependence.  In the Landau-like version assume that the
charges stop instantaneously,
remain at rest for a time $\tau$, then instantly accelerate to
final velocities $\pm v_f$.  This intensity distribution is:
\begin{eqnarray}
\frac{d^2I}{d\omega d\Omega} &=& \frac{Z^2 \alpha}{\pi^2}
\sin^2\theta \cos^2\theta \left\{ \frac{[v_0^2-v_f^2\cos\omega\tau
-v_0^2 v_f^2 (1-\cos\omega\tau)\cos^2\theta]^2}
{(1-v_0^2\cos^2\theta)^2 (1-v_f^2\cos^2\theta)^2} \right. \nonumber \\
&+& \left. \frac{v_f^4 \sin^2\omega\tau}{(1-v_f^2\cos^2\theta)^2}
\right\} \, .
\end{eqnarray}
The period of oscillation is just the time delay between stopping
and sudden acceleration.  This is essentially what happens in the
Landau-like model, modified by the smearing which follows from
a continuous acceleration rather than a time delayed one, and by
the relativistic effect in the phase factor as mentioned before.

\section{Conclusion}

In this paper we have studied two very different and extreme models
of the space-time evolution of the charges in
central nucleus-nucleus collisions
at a RHIC energy of 100 GeV per nucleon in the center of mass frame.
The models were constructed to yield the same final rapidity
distribution of the net outgoing electric charge.  This means that
measurement of hadron spectra alone cannot discern them.  Very low
energy bremsstrahlung cannot discern them either because it depends
only on the difference between the incoming and outgoing currents,
which are identical.  Bremsstrahlung photons of higher energy can
distinguish these extreme models.  For the Landau-like model
photons of energy 100 MeV or so is quite sufficient to infer
the time scale for expansion.  Bremsstrahlung photons with
energy up to 200 MeV may be able to infer the deceleration time
in Bjorken-like models but one needs to look at angles much
greater than where the peak of the spectrum occurs, which is 1 degree.
Whether it is really possible to design a detector that can separate other
sources of photons
of this energy is an open question, but at least we have shown
that the physics to be learned is interesting, perhaps justifying
an experimental effort \cite{Jack}.

\section*{Acknowledgements}
We greatly appreciate conversations with J. Sandweiss.
This work was supported by the U.S. Department of Energy under grant
DE-FG02-87ER40328.

\end{document}